# Cyber Threat Detection and Vulnerability Assessment System using Generative AI and Large Language Model

1st Keerthi Kumar. M
Department of Electronics and Communication Engineering
Nitte Meenakshi Institute of Technology, Nitte (Deemed to be University)
Bengaluru, India
keerthi.kumar@nmit.ac.in

2nd Swarun Kumar Joginpelly
Independent Researcher
USA
swarunjogi@gmail.com

3rd Sunil Khemka
Persistent Systems
Chicago, IL, USA
sunilkhemka.tech@gmail.com

4th Lakshmi. S R
Department of CS (DS)
Dayananda Sagar Academy of Technology and Management
Bengaluru, India
lakshmi-csds@dsatm.edu.in

5th Navin Chhibber
Infinity Tech Group
Sunnyvale, CA, USA
naveenchibber.research@gmail.com

*Abstract*— Background: Cyber-attacks have evolved rapidly in recent years, many individuals and business owners have been affected by cyber-attacks in various ways. Cyber-attacks include various threats such as ransomware, malware, phishing, and Denial of Service (DoS)-related attacks. Challenges: Traditional models such as Generative Artificial Intelligence (AI) and Security Bidirectional Encoder Representations from Transformers (BERT) were implemented to detect cyber threats. However, the existing Security BERT model has a limited contextual understanding of text data, which has less impact on detecting cyber-attacks. Proposed Methodology: To overcome the above-mentioned challenges, Robustly Optimized Bidirectional Encoder Representations from Transformers Pretraining Approach (RoBERTa) model is proposed which consists of diverse words of vocabulary understanding. Initially, data are extracted from a Packet Capture (PCAP) file and encrypted using Fully Harmonic Encryption (FHE). Subsequently, a Byte-level and Byte Pair Encoding (BBPE) tokenizer was used to generate tokens and help maintain the vocabulary for the encrypted values. Then, these values are applied to the RoBERTa model of the transformer with extensive training. Finally, Softmax is used for the detection and classification of attacks. The proposed RoBERTa model achieved better results than the existing BERT model in terms of accuracy (0.99), recall (0.91), and precision (0.89) respectively.

## I. INTRODUCTION

In the modern era of technology, cybersecurity has evolved gradually and is mostly used by businesses, governments, and individuals around the world. The vital role of cyber security played in protecting computer systems from vulnerable attacks such as phishing, ransomware, Distributed Denial of Service (DDoS) and malware-related attacks [1]. This can be achieved by utilizing the Vulnerability Assessment System (VAS), which is a systematic process that helps in determining the potential risks and reports to the user before an attack occurs, which helps the users from the security attacks before they occur [2]. However, as technology is evolving rapidly in terms of cloud computing, Internet of Things (IoT), and the Artificial Intelligence (AI) personal data is being shared on the Internet, which creates a challenging task for users to secure data [3]. There are a few attacks that use AI to assist attacks, such as deep fake; it generates realistic images, audio, text, and videos related to user feed. Furthermore, the generated data can be used for blackmailing, faking biometric authentication, and so on [4]. Moreover, cloud computing introduces cyber threats and vulnerabilities such as insecure API, unauthorized access, and data breaches. To address the above-mentioned challenges, such as deep fake, data breaches, and faking biometric authentication, Generative Ai (GAI) and Large Language Model (LLM) are utilized [5]. Further, autoencoders were implemented, which helped in focusing on the important data and in improving the detection of cyber threats [6]. In addition, AI and LLM been used to identify vast unknown threats, including phishing and malware. Moreover, (VAS) was utilized to identify weak data and check for vulnerabilities of the system before attackers accessed the data [7]. State-of-the-Methods includes the ChatPhishDetector, which is used to detect various phishing sites using a Large Language Model (LLMs). Moreover, the proposed system utilizes a web crawler to access various input URL and collect data from the pages [8]. However, some LLMs were trained for a certain period of time rather than being up-to-date, which made it difficult to detect phishing and non-phishing sites. Similarly, utilized the hybrid Grasshoper-Crowsearch Optimization (GSCSO) technique with an Isolated Heuristic Neural Network (IHNN) technique used to predicted the flow of data as intrusive or normal [9]. Furthermore, Iterative Principal Component Analysis (IPCA) is used to eliminate unwanted data, which helps detect intrusions more accurately. However, the proposed neural network simply classifies the threat as benign or malicious instead of detecting a specific threat [10].

The vital-contributions of this research are described as follows,

- The Fully Harmonic Encryption (FHE) was used for encrypting numerical values inorder to understand textual representation and Byte-level and Byte Pair Encoding (BBPE) tokenizer from hugging face transformer used to encode and decode text for getting original meaning.

- In securityBERT+, data goes through padding for the same length of values for stable training, and preprocess the input data through text normalization, tokenization and frequency filtering for data cleaning.
- Robustly Optimized Bidirectional Encoder Representations from Transformers Pretraining Approach (RoBERTa) model improves the accuracy of cyber threat detection through word embedding, multi-head attention, and feed-forward for extensive training.

The remainder of this paper is organized as follows: Section 2 explains the literature, Section 3 describes the proposed methodology with a pictorial presentation, Section 4 describes the results of the proposed methodology, and further corresponding discussion is presented in Section 5. Finally, the conclusions of this study are presented in section 6.

## II. Literature Review

Ferrag et al. [11] developed a Privacy-Preserving Bidirectional Encoder Representation from Transformer (BERT)-Based Lightweight Model for cyberthreat detection by using LLMs. Initially data was collected from Industrial Internet of Things (IIoT) dataset, which contains 15 attacks related to cyber threat. For feature extraction, Packet Capture (PCAP) was utilized to extracts data from a dataset and further stored in separate file. Privacy preserving fixed-length encoding (PPFLE) is used for hashing the features, and a BBPE tokenizer forms the sentence using character compression and rearrangement. Subsequently, a BERT takes the input ID and mask attention values and provides BERT. BERT uses self-attention to understand attacks and provides the final output using Softmax. This model was efficient in detecting attacks. However, PPFLE was a weak technique that can be used only by hackers.

Shestov et al. [12] developed Vulnerability Detection using fine-tuned Large-Language Models (LLM). The data collection consists of Manually-Curated, Common Vulnerabilities and Exposures (CVEfixes) and VCMatch, which was open source and fix commits, Vulnerabilities of database and fix commits respectively. During data pre-processing, the cleanup and functionality of irrelevant changes was fixed in the modified dataset. In this methodology, LLM was selected for secure detection and used. The LLM used two primary techniques: next-token prediction (NTP) and Binary Classification (BC). In NTP code, the input sequence was packed one after another, and BC also generated for prediction purposes. The BC label was used for vulnerability values of 0 or 1, and cross-entropy was calculated for probability prediction. The advantage of this model was that it detected vulnerabilities efficiently. However, the model required modifications to the nontrivial for the accumulation of the gradient.

Purba et al. [13] developed Large Language Models (LLM) using Software Vulnerability Detection. Data collection consists of Common Vulnerabilities and Exposures (CVEfixes) and code gadgets, which have snippets of c++ and c codes with an overflow of buffer vulnerabilities. In the feature extraction, the snippet and code slices were extracted, and code comments were deleted. In methodology 16 attentions and 28 layers of combinations like batch size and epoch. Subsequently, source codes were analyzed using pattern matching so that the LLM learned the patterns of code using the program technique. For Structured Query Language (SQL) injection, LLM detects using the variable of string and concatenation of string and detects the exact location. The advantage of this model was that the combination of the program technique and LLM achieved better performance in vulnerability detection. However, this model was not efficient in identifying sanitized data.

Heiding et al. [14] developed Large Language Models (LLM) using Devising and Detecting Phishing Emails. The data collection consisted of custom-made emails that consists of 725k emails are of legit and phishing emails. Phishing mail was classified into four categories, with the first being the control group (CG). In the CG, some phishing emails generated by different LLM were tested with the generative pre-trained transformer (ChatGPT) and answered accurately. This generation of LLM was the second category of phishing emails. Third, the V-TRAID of personalization adds content and features, such as logo and unsubscribe, to the email. Fourth, V-TRAID and Generative Pre-trained Transformers (GPT) of personalization tested mail using Bard and Claude with the exclusion of graphical content. The advantage of proposed model was it effectively detected phishing emails. However, the model exhibited low deviation in the phishing context.

Saha et al. [15] developed a Paradigm Shift using LLM for Security operations center (SoC) security. The data collection consisted of 156 problems, included function codes, automatic testing, and Verilog tasks. In this methodology, the vulnerabilities of the security-exit condition, encoding state, and deadlock were considered. Next, untheorized access was detected by the Generative Pre-trained Transformer (GPT) of 4 and 3.5 respectively, and tested using the Advanced Encryption Standard (AES) design. For this assessment, utilized two techniques such as Contextual Test (CT) and context-free text (CFT) for techniques of insertion in AES designs, and temperature settings from 0 to 1. These coding issues were also detected using the GPT-4 and GPT-3.5. The advantage of the proposed model was that it revolutionizes the reasoning and coding tasks. However, this model was a limited vulnerability.

## III. Methodology

This study proposes an RoBERTa deep learning-based model designed to efficiently process and detect various cyber threats. The methodology begins with the collection of network-related data from the edge IIoT dataset, which comprises raw network-related traffic data in PCAP format. Furthermore, to prepare the data, a few pre-processing steps were applied, including text normalization to ensure readability, tokenization to divide the input data into small chunks, frequency filtering to remove useless tokens, and vocabulary creation to build a structured representation. Feature extraction is then performed using the PCAP file and full harmonic encryption to collect the main important featured data from the PCAP log file. Additionally, the BBPE technique was employed to encode the collected data to ensure privacy. Finally, the extracted features were fed into the proposed RoBERTa model to improve the context and robust training for learning patterns, classifying threats, and detecting anomalies. The architecture of the methodology is illustrated in below Figure 1.

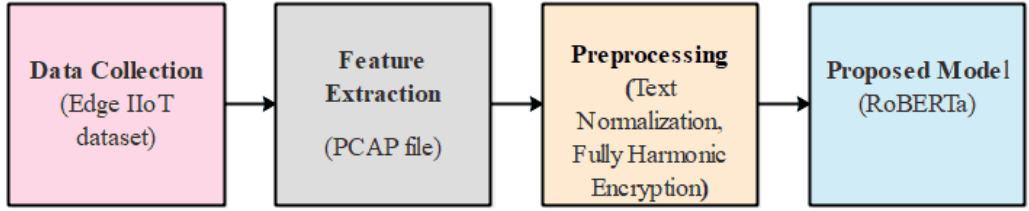

Fig. 1. Workflow of proposed methodology

### A. Dataset

Initially, input data were collected from the Edge Industrial Internet of Things (IIoT) dataset [11], which consists of real-life traffic data utilized for detecting several types of security threats in IoT and IIoT-based applications. This dataset comprises 15 different types of security attacks, which are further grouped into five major threat categories: Denial of Service (DoS), man-in-the-middle (MITM), information gathering, malware and injection attacks. Similarly, benign traffic data were used for the multiclass classification. Moreover, in each category, the attacks were classified into the following types: Hypertext Transfer Protocol (HTTP), User Datagram Protocol (UDP), Transmission Control Protocol synchronize (TCP) SYN and Internet Control Message Protocol (ICMP) flood. Gathering of information attacks include fingerprinting of OS, ports, and vulnerability scanning. Moreover, MITM includes a Domain Name System (DNS) spoofing attack and an Address Resolution Protocol (ARP) spoofing attack. Further, injection attacks include SQL Injection, uploading, and cross-site scripting (XSS) attacks. Also, malware attacks include ransomware attacks, backdoors, and password cracker-related attacks.

### B. Feature Extraction

Dataset features were extracted using a Packet Capture (PCAP) file to ensure the efficiency of the exact features. The PCAP file contains malicious attacks with timestamps, data packets, and headers, with 61 diverse and distinct features. Some key features include $arp.hw\_size$, $ip.src\_host$, and $ip.dst\_host$ of the data consisting of character strings and unsigned integers. These features contain various types of attacks that are sufficient for the analysis. Subsequently, the PCAP file feature values were written as Comma-Separated Values (CSV) for training.

In fact, features contain binary, nonbinary, and categorical formats of numerical representations, which makes it difficult to understand the RoBERTa model. Proposed model was specifically designed to understand the relationship between words. Therefore, the numerical and binary representations are transformed into textual representations using the Fully Harmonic Encryption (FHE). FHE performs encryption and generates a sequence suitable for the Roberta model for training. FH9E works using public and private keys, and in encryption, the numbers are converted into ciphertext by shifting the positions. The public key is used to encode the numerical values, and the private key receives encrypted data. The formula for the cipher text is shown in Equation (1) respectively:

$$C = (p + k) \, mod \, 26 \quad (1)$$

Where $C$ is the ciphertext, $p$ is the numerical value, $k$ is the shifting position, and mod is the modulo function.

Here, some of the features were removed to eliminate noise data such as time related, video or audio frames data were removed during the training due to different IP addresses of host and source leads to overfitting. Further, these extracted features are fed as input into Byte-level and Byte Pair Encoding ( ) for tokenization, which is discussed in following section.

*1) Tokenization:* Using FHE, numerical representations are converted to textual representations related to Natural Language Processing (NLP). Combining all features and performing training are the semantics of destroying attacks. Here, utilized BBPE Tokenizer from the Hugging-Face Transformers Library, which divides sentences into words known as tokens. In this BBPE tokenizer, the decoder first uses the private key of decryption, analyzes the sentence, and converts it into bytes using Unicode Transformation Format (UTF)-8 encoding. In the decoder process, bytes are rearranged and words are formed using out-of-vocabulary (OOV). This language was not understood by humans, so the OOV broke down this language and processed the traffic data effectively.

### C. Pre-Training of Security BERT+

*1) Data pre-processing:* In this section, the extracted data from BBPE is used as input data it was a chunk size of 5000 respectively. therefore, the chunk size data were divided by the length eval_data by chunk_size. This information is stored in the $input\_id\_chunk$ and $attention\_mask\_chunk$ for pre-processing. Subsequently, there were different lengths representing the data, which was difficult for the model to evaluate. For different lengths, padding was performed to represent the same length, and attention_mask_chunk was used to mask padding to avoid confusion. Input_id tells the 0 for padding token and 1 for the real token.

*2) Text Normalization:* Text Normalization involves transforming the input text into lowercase, removing punctuation and lemmatization, and stemming to reduce the base form. This process is crucial to RoBERTa's training. The formula for normalization is given in Equation (2):

$$n(D) = \{n(d) | d \in D\} \quad (2)$$

Where, $n(D)$ is normalization, d is a document

*3) Tokenization:* In tokenization, text is broken down into subwords. These tokens are the basic texts of units that are important for model training. The tokenization formula is given in Equation (3):

$$t(d) = \{t(w) | w \in d, d \in D\} \quad (3)$$

Where $t(d)$ is the tokenization, $w$ is the word of the token, $D$ is a set.

*4) Frequency Filtering:* A frequency filter was used to remove rare words that occurred in the text, which did not provide much information in the text. The frequency-filtering formula is shown in Equation (4):

$$f(D,F) = \{w \in D | freq(w,D) \geq F\} \quad (4)$$

Where, $f(D,F)$ is a high pass filter, $freq(w,D)$ is a minimum frequency

*5) Vocabulary Creation:* It creates vocabulary for incorrect and confused sentences, and special token addition performs the padding, starting, and ending of sentences. Equations (5) and (6) show the formulas for vocabulary creation and special token shown in Equation (5 and 6):

$$v(D,V) = \{w | w \in D, rank(w,D) \leq V\} \quad (5)$$

$$v' = v \cup S \quad (6)$$

Where $v(D,V)$ is the vocabulary, $w$ is the word, $D$ is the document, $v$ is the vocabulary, and $S$ is a special token. Further, the output will be a custom vocabulary which is built on mostly used tokens and fed to the proposed model as input.

*D. Training with RoBERTa Model*

In this section, proposed RoBERTa model is trained on the input data where four encoder layers are utilized, and dropout and classifier layers are involved. Furthermore, the model contains the embedding layer, self-attention, a normalization layer, masked language modelling, a feed-forward neural network, and a softmax layer. The main advantage of proposed RoBERTa model is it utilized PPFLE technique for protecting sensitive data. Moreover, BBPE technique leveraged the hashing code for ensuring efficient encoding of collected data.

First, word embedding was used to convert text representation into a numerical representation using word2vec, which was developed by Google. Word2vec uses mathematical representation based on the training of neural networks, and data consists of Internet sources of corpus text. where Word2vec represents the dimensionality of the vector. Subsequently, positional encoding (PE) was performed for the context representation. In PE, the odd value was calculated using the sine function, and the even row was calculated using the cosine function, which is represented in Equations (7)-(8):

$$PE_{(Pos,2i)} = sin(\frac{pos}{10000^{\frac{2i}{d_{model}}}}) \quad (7)$$

$$PE_{(Pos,2i+1)} = cos(\frac{pos}{10000^{\frac{2i}{d_{model}}}}) \quad (8)$$

Here, $PE$ is the positional encoding of odd and even dimensions, $pos$ is the token of the position sequence, $i$ is the dimensional vector of the positional encoder, $d_{model}$ is the embedding space of the model.

These input values were parallel to the encoder for multi-head attention using a manufacturing strategy. In multi-head attention, all words form vectors one by one and multiple with a query, key, and value, respectively. First, the input value is multiplied by the weight matrix of the query and the query matrix is given. Similarly, key and value multiply by the weight matrices of key and value, and give the output of key and value, respectively, using masked language modelling dynamically. Subsequently, Softmax multiples the query and key by dividing the dimensionality of vector embedding. The Softmax formula is given by Equation (9).

$$Attention\ weights = softmax(\frac{QK^T}{\sqrt{d_K}}) \quad (9)$$

From the above Equation, $Q$ denotes the query, $K^T$ is the transpose of the key matrix, $d$ is the dimensionality of the matrix.

The multi-head attention values were normalized, where all values were normalized with a mean of 0 and a variance of 1. Subsequently, it feeds forward a neural network by using a Gaussian Error Linear Unit (GELU) function. The key advantage of using this activation function is that, even if the input values are negative, it provides training to the values, unlike the RELU function, which eliminates the input values when it approaches zero. The GELU activation function provided more training and accurate results. The dropout layer is used to increase the performance by deleting some neurons that provide less information.

Then, the values for the normalization and again go to the second encoder and repeat the process. The values are sent to the four encoder layers for stable training and accurate detection. Finally, the softmax function detects and classifies 15 attacks, of which 14 are malicious and one is normal. Moreover, further section describes about the detailed experimental results of the proposed model.

IV. EXPERIMENTAL RESULTS

Experimental results were obtained in a high-performance computing environment to ensure efficient model training and validation. This environment uses multicore processors such as Intel i7, at least 16 GB of random-access memory (RAM), and a GPU such as NVIDIA GTX 1080 to perform DL training. In particular, the software environment includes Python and DL libraries such as TensorFlow and PyTorch. The proposed RoBERTa model results in an optimized, highly relevant input set that enhances accuracy and reduces redundancy. The evaluation metrics utilized are formulated as follows in Equations (10)-(12):

$$Accuracy = \frac{TP+TN}{TP+TN+FP+FN} \quad (10)$$

$$Precision = \frac{TP}{TP+FP} \quad (11)$$

$$Recall = \frac{TP}{TP+FN} \quad (12)$$

In the above Equation, TN and TP signify true negatives and positives, respectively, and FN, FP represent false negatives and positives, respectively.

*A. Performance analysis*

The performance of the proposed RoBERTa model was compared with those of various existing models, such as ChatGPT 3.5-turbo, Falcon, and Alpaca-LoRA, as described in the Table 1, with their respective metrics include accuracy, precision and recall.

TABLE I. PERFORMANCE ANALYSIS OF PROPOSED ROBERTA MODEL

| Performance models | Accuracy | Precision | Recall |
|---|---|---|---|
| ChatGPT 3.5- turbo (16k context) | 0.90 | 0.95 | 0.82 |
| Falcon | 0.92 | 0.81 | 0.85 |
| Alpaca-LoRA | 0.94 | 0.89 | 0.79 |
| Proposed RoBERTa | 0.99 | 0.91 | 0.89 |

From Table 1, the proposed RoBERTa-based transformer model provides better results, with an accuracy of (0.99), a precision of (0.91) and a recall of (0.89) respectively. Compared with other existing models, such as ChatGPT 3.5-turbo, it achieved an accuracy of (0.90) with very high precision (0.95) but slightly lower recall (0.82). Falcon obtained an accuracy of (0.92), a precision (0.81) and a recall of (0.85). Alpaca-LoRA demonstrated higher accuracy (0.94) and precision (0.89) than all existing models, although its recall (0.79). Overall, the proposed RoBERTa model significantly outperformed the other models, offering the most reliable and robust results across all the three-evaluation metrics.

*B. Comparative Analysis*

The proposed RoBERTa model was compared with other existing models, such as Security BERT [11], and the existing model was used to detect security threats from various sources. The proposed model was compared with the Security BERT [11] model in the following Table 2 respectively.

TABLE II. COMPARATIVE ANALYSIS OF PROPOSED ROBERTA MODEL WITH EXISTING MODEL

| Performed Models | Accuracy | Precision | Recall |
|---|---|---|---|
| Security BERT [11] | 0.98 | 0.87 | 0.84 |
| Proposed model (RoBERTa) | 0.99 | 0.91 | 0.89 |

The table above shows a detailed analysis of the proposed model, along with its existing model and its accuracy. The proposed model provided efficient and better results, with an accuracy of 0.99, precision of 0.91, and recall of 0.89. However, the existing Security BERT [11] model achieved an accuracy of 0.98, precision of 0.87, and recall of 0.84. Although the proposed RoBERTa model utilizes the Packet Capture (PCAP) log file, which is used to collect data from the threat whenever a threat occurs, it captures data related to threats, such as the source address, destination, packet size, duration, and length of the data. Moreover, it collected 61 different features related to threats and detected specific threats related to the information captured. Therefore, the proposed RoBERTa model attained better accuracy than Security BERT.

## V. DISCUSSION

In this paper, presented a RoBERTa transformer-based model for the detection of various threats and utilized the FHE technique to encode the collected data into a privacy-preserving format, such as a hashing format. The RoBERTa model was compared with several other LLM models such as ChatGPT 3.5 turbo, falcon, and Alpaca-LoRA. However, the ChatGPT 3.5 turbo faces several challenges while detecting security threats, some of which include inconsistent responses and a high rate of false positives. The Falcon model suffers from a lack of real-time knowledge, owing to its static knowledge. The Alpaca-LoRA requires powerful high-end GPUs to process data related to threats. Moreover, the Alpaca-LoRA model does not understand the contextual meaning or domain-specific context present in the cyber data. Although the proposed RoBERTa model excelled in the detection of cyber threats by utilizing the PCAP log file as a feature extraction process, it stored only the required cyber data in a CSV file for further processing.

## VI. CONCLUSION

In this study, the RoBERTa model is proposed for detecting various types of threats. The detection process begins by collecting cyberthreat data from the PCAP file. This PCAP file consists of the information log of incoming data, such as the source destination address, IP address, and size and duration of the threat data collected in the PCAP file. Furthermore, feature extraction is performed by collecting the most relevant details of cyber threats from the PCAP log file and storing them in a CSV file for further processing. The FHE technique was employed to convert the collected data into a privacy format, such as hashing, and concatenate all the collected data to encrypt the data for privacy. Subsequently, tokenization was applied to the collected data using byte-level encoding (BBPE). Furthermore, each token is assigned a distinct ID, and the tokens with their respective id are sent to the RoBERTa model for embedding. Multi-head attention was then utilized to capture the relationship between tokens and to detect the type of threat. The result of the proposed RoBERTa outperforms the current Security BERT model in terms of accuracy of (0.99), precision (0.91) and recall (0.89), when compared with existing Security BERT model. In the future, this research will be extended by finetuning the RoBERTa model on domain specific datasets which helps in detecting various types of cyber threats in various domains. Moreover, integrating a light weight transformer model will be beneficial for deploying in real time environments.